\documentclass{article}\usepackage{amssymb}

		\newcommand{\ec}[1]{(\ref{#1})}

   	\def\barr{\begin{eqnarray}}\def\earr{\end{eqnarray}}
   
   	\newcommand{\parent}[1]{\left(#1\right)}
   	\newcommand{\corchet}[1]{\left[#1\right]}
   	\newcommand{\llaves}[1]{\left\{#1\right\}}
         
\begin{document}
\fontsize{10pt}{13pt}\selectfont

\begin{titlepage}
\title{$D=11$ Supermembrane wrapped on calibrated submanifolds}
\author{J. Bellor\'{\i}n$^1$\footnote{\tt jorge.bellorin@uam.es} 
\hspace{3mm} and \hspace{3mm} A. Restuccia$^2$\footnote{\tt arestu@usb.ve} \\ {} \\
{\small{\it $^1$ Instituto de F\'{\i}sica Te\'{o}rica UAM/CSIC, Facultad de Ciencias C-XVI,}} \\ {\small {\it C. U. Cantoblanco, E-28049 Madrid, Spain.}} \\ {\small {\it $^2$ Departamento de F\'{\i}sica, Universidad Sim\'on Bol\'{\i}var,}} \\ {\small {\it 1080-A Caracas, Venezuela.}} \\ {} \\ {\small October, 2005 }}
\date{}
\maketitle

\begin{abstract}
{We construct the Hamiltonian of the $D=11$ Supermembrane  with topological conditions on configuration space. It may be interpreted as a supermembrane theory where all configurations are wrapped in an irreducible way on a calibrated submanifold of a compact sector of the target space. We prove that the spectrum of its Hamiltonian is discrete with finite multiplicity. The construction is explicitly perfomed for a compact sector of the target space being a $2g$ dimensional  flat torus and the base manifold of the Supermembrane a genus $g$ compact Riemann surface. The topological conditions on configuration space work in such a way that the $g=2$ case may be interpreted as the intersection of two $D=11$ Supermembranes  over $g=1$ surfaces, with their corresponding topological conditions. The discreteness of the spectrum is preserved by the intersection procedure. Between the configurations satisfying the topological conditions there are minimal configurations which describe minimal immersions from the base manifold to the compact sector of the target space. They allow to map the $D=11$ Supermembrane with topological conditions to a symplectic noncommutative Yang-Mills theory. We analyze geometrical properties of these configurations in the context of Supermembranes and  D-branes theories. We show that this class of  configurations also minimizes the Hamiltonian of  D-branes theories.}
\end{abstract}
\thispagestyle{empty}
\end{titlepage}

\newpage\section{Introduction}
Recently, the $D=11$ Supermembrane\hyphenation{Su-per-mem-bra-ne} \cite{Bergshoeff}, under a specific topological restriction on configuration space, was shown to have a discrete spectrum \cite{LB+MG+IM+AR,LB+MG+AR}, see also \cite{Duff,IM+AR+RT,IM+JO+AR,MG+AR}. The topological restriction in \cite{LB+MG+IM+AR,LB+MG+AR} consists in the irreducible wrapping of the supermembrane on a sector of the target space compactified in $S^1\times S^1$, which induces a nontrivial central charge. In distinction, but not in contradiction, the spectrum of the $D=11$ Supermembrane, when the target space is $D=11$ Minkowski space, was shown, in a remarkable work, to be continuous from zero to infinity \cite{deWitt+Luscher+Nicolai,deWit+M+N}; even if the target space  has a compactified sector it was argued in \cite{deWitt+Peeters+Plefka} that the spectrum should still be continuous. It is the topological restriction on configuration space that changes the structure of the interacting potential in the sense that it eliminates the string-like spikes responsible, in supersymmetric models, of having a non bounded resolvent for the Hamiltonian. In all these arguments a reguralized  Hamiltonian was used, the $N$ going to infinity limit was not addressed and it remains as an open problem. 

In this paper we study the generalization of the topological restrictions to be imposed on the configuration space of a $D=11$ Supermembrane in order to obtain a discrete spectrum. It turns out that the topological restrictions can be presented in terms of
an irreducible wrapping of the supermembrane on a calibrated submanifold of a compactified sector of the target space. The topological restrictions are also analogous to a flux condition which has been used to stabilize the moduli in phenomenological models of superstrings \cite{MPG,JC+MPG+FQ+AU}. Nontrivial central charges in the supersymmetric algebra of the supermembrane are also induced by the wrapping of the supermembrane on the calibrated submanifold. From a geometrical point of view this topological condition means that we are considering the quantization of a field theory on a nontrivial principle bundle and an associated vector bundle. In \cite{LB+MG+IM+AR,LB+MG+AR} the base manifold  was a $g=1$ compact Riemann surface, a torus, and the target space Minkowski times $S^1\times S^1$. In this work we take the Supermembrane with base manifold a genus $g\geq 1$ compact Riemann surface and target space with a compact sector being a $2g$ dimensional flat torus. It becomes natural, although technically involved, from the construction we will present how to extend the formulation to more general flat compact sectors of the target space. 

Within the configurations satisfying the topological restrictions there exists one which minimizes the Hamiltonian, the minimal immersion. In fact, it is the minimal immersion that defines the calibrated submanifold of the compactified target space on which the supermembrane wraps. The minimal immersion allows to map the original formulation of the supermembrane restricted by the topological conditions to a simplectic noncommutative Yang-Mills theory formulated over the calibrated submanifold. The existence of the minimal immersion was implicit in the construction in \cite{LB+MG+IM+AR,LB+MG+AR}. The final goal will be to show that the spectrum of these new models of the supermembrane is discrete. Since the existence of the minimal immersion of the base manifold into a submanifold of the compact sector of the target space is crucial in the construction, it imposes a non trivial restriction on the supermembranes models for which its reguralized Hamiltonian has a discrete spectrum, at least with the topological conditions we consider.  

We also study general properties of the minimal immersions, specifically in relation to the dynamics of D$p$-branes arising from the Dirac-Born-Infeld action formulated over topologically non trivial manifolds. We will show that a minimal immersion of a K\"ahler manifold of dimension $p$ into the target space together with the gauge field whose curvature is the K\"ahler two form always provide a regular solution of the DBI field equations. Moreover they are minima of its corresponding Hamiltonian. We will first discuss these geometrical properties of the minimal immersions and then proceed to the construction of the new supermembrane models, with topological restrictions, where they play a relevant role.

The geometry of supersymmetry preserving embeddings of branes may be
understood in terms of calibrated geometries \cite{Harvey+Lawson,Harvey,Cascales+Uranga}. They allow to determine the supersymmetric cycles on which the membrane can wrap in some holonomy manifolds. There is a closed relation between calibrated submanifolds and intersecting branes, for a review see \cite{Smith}. In particular we will construct a supermembrane vacuum, with topological restrictions, over a genus $g=2$  compact Riemann surface which is exactly the intersection of two supermembrane vacua, with their corresponding topological restrictions, over $g=1$ compact Riemann surfaces. The latest preserve half of the supersymmetry while the former preserve a quarter of the original supersymmetry. The topological restrictions work in such a way that the gluing of the two $g=1$ surfaces, to give rise to a $g=2$ surface, leads to the intersecting supermembrane theory. We remark that we are considering here a complete theory over a compact surface, not just a solution of $D=11$ Supergravity. We will show that the resulting intersecting supermembrane theory also has a discrete spectrum. The extension of the construction and of its interpretation for $g>2$  compact Riemann surfaces follows on the same lines.

This paper is organized as follows: In section \ref{Solutions} we discuss geometrical properties of the minimal immersions. In section \ref{Membrane} we analyze the $D=11$ Supermembrane, we first define its configuration space by imposing the topological conditions in subsection 3.1. In subsection 3.2 minimal immersions are explicitly constructed in terms of the holomorphic one-forms on the Riemann surface and in the following subsection the unbroken supersymmetries of these configurations is calculated. In subsection \ref{SymplecticYangMills} we reformulate the supermembrane theory in terms of a symplectic Yang-Mills theory and in 3.5 we show that its spectrum is discrete.

\section{Minimal configurations of D$p$-branes}\label{Solutions}
In this section we study geometrical properties of minimal configurations, that is configurations minimizing the world-volume. In particular of a class of regular minimal configurations which will be relevant in the quantization of supermembranes with topological restrictions, and in general in the quantization of D-branes. We first will show that a regular configuration constructed by taking any minimal immersion from a K\"ahler manifold into a target space  together with the connection, on a non- trivial $U(1)$ principle bundle with base manifold the K\"ahler one,  whose curvature is the K\"ahler two-form solves the D-brane field equations. These configurations minimizes the Dirac-Born-Infeld Hamiltonian over the K\"ahler manifold. It was shown in \cite{B+R} that any connection of this kind on a non-trivial $U(1)$ principle bundle  over a K\"ahler manifold exactly solves the Born-Infeld field equations and that they are minima of the BI Hamiltonian. It is then natural to extend them to solutions of the DBI theories. This geometrical picture exactly  appears in the $D=11$ Supermembrane with topological restrictions, it is directly related to the topological conditions we are going to impose on the configuration space of the supermembrane. 

The extended self-dual configurations constructed in \cite{B+R} were defined on Riemannian manifolds of even dimension $2k$ where a curvature two-form satisfy the conditions 
\begin{equation}\label{extendedselfdual} 
	^\star P_m \sim P_{k-m} 
\end{equation} 
for all possible $m$, where 
\begin{equation}
	P_m \equiv F\wedge F\wedge\ldots\wedge F 
\end{equation}
($m$ factors) and $^\star$ denotes the
Hodge dual. That is, the set $\llaves{P_m}$ is left invariant by the Hodge
dual operation. In four dimensions it reduces to $^\star\parent{F\wedge
F}=constant$ and $^\star F\sim F$, the well-known self-duality condition which
solves the Maxwell equations. In two dimensions $^\star F=constant$ represents
a magnetic monopole configuration.

It was shown in \cite{B+R} that those configurations, whenever they exist, solve the field equations of the Born-Infeld Theory. This theory is
described by the following abelian gauge action over a pseudo-Riemannian
manifold $\Sigma$ \begin{equation}\label{BI} S_{BI}[A] = \int\limits_\Sigma
d^{p+1}\sigma \sqrt{-\det\parent{G_{\mu\nu} + F_{\mu\nu}}} \end{equation} where
$G_{\mu\nu}$ is a given metric over $\Sigma$. The extended self-dual
configurations, which are non-trivial regular configurations over the spatial part of $\Sigma$, are static solutions of the field equations.
Moreover, in \cite{B+R} it was proven they are minima of the corresponding
Born-Infeld Hamiltonian.

A non trivial example of extended self-dual configurations arises from
K\"ahler manifolds of non trivial topology. The K\"ahler form verifies the extended self-duality condition (\ref{extendedselfdual}) if the metric is precisely the K\"ahler metric. One
may then introduce a connection on a $U(1)$ principle bundle over the
K\"ahler manifold endowed with a collection of local sections, such that
the pullback of its curvature correspond to the K\"ahler form. The local
pullbacks of the connection may be taken as gauge fields solving the field
equations of \ec{BI}, $G_{\mu\nu}$ being the K\"ahler metric. We remark
that the gauge fields are non-trivial as consequence of the non-trivial
topology of the bundle, the latest implies in particular the non existence
of a global section. Each gauge field is regular on the open set where it
is defined and two overlapping fields are related by a $U(1)$ gauge
transformation, as it is required in order to have a globally well defined
connection. Particular examples of connections on principle bundles whose
curvature yields the K\"ahler form of the base manifold are the canonical
connections given by the Hopf fibering $S^{2n+1}\rightarrow \mathbb{C}P^n$
presented by Trautman \cite{Trautman} as solutions to the Maxwell
equations, the lowest dimensional case $S^{3}\rightarrow \mathbb{C}P^1$
corresponding to a magnetic monopole.

The low energy effective dynamics of a D-brane is described by the
Dirac-Born-Infeld action \cite{Leigh}, which possesses the same structure
of the Born-Infeld one but leaving the metric to be a dynamical object
induced on the D-brane world-volume by scalar fields. The existence of non-trivial regular solutions of the Born-Infeld theory over K\"ahler
manifolds is a strong motivation to consider the problem of minimal
configurations of D-branes on K\"ahler manifolds of non trivial topology.

The Dirac-Born-Infeld action of a D-brane is a functional of two fields
over $\Sigma$; they are the set of components of a map $\Sigma \rightarrow
{\cal T}$, where ${\cal T}$ is the target space, describing the embedding
of $\Sigma$ into ${\cal T}$, and an $U(1)$ gauge field $A$. The action is
\begin{equation}\label{covactionA}
  S\corchet{X,A} = \int\limits_\Sigma d^{p+1}\sigma\sqrt{-\det
  E_{\hat{\mu}\hat{\nu}}}
\end{equation}
where
\begin{eqnarray}
&&  E_{\hat{\mu}\hat{\nu}} = G_{\hat{\mu}\hat{\nu}} +
    F_{\hat{\mu}\hat{\nu}}, \\
&&  G_{\hat{\mu}\hat{\nu}}
    =\partial_{\hat{\mu}}X^{\hat{M}}\partial_{\hat{\nu}}X^{\hat{N}}
    g_{\hat{M}\hat{N}}, \\
&&  F_{\hat{\mu}\hat{\nu}}=
    \partial_{\hat{\mu}}A_{\hat{\nu}}-\partial_{\hat{\nu}}A_{\hat{\mu}},
\end{eqnarray}
\[ \hat{M}=0,M ,\hspace{4mm} M=1,\ldots,\dim {\cal T}-1,\hspace{4mm}
\hat{\mu}=0,\mu ,\hspace{4mm} \mu=1,\ldots,p.
\] $g_{\hat{M}\hat{N}}$ denotes a fixed metric on ${\cal T}$
and $X^{\hat{M}}$ are the components of the map.

The field equations derived from \ec{covactionA} are
\begin{eqnarray}
&& 	\parent{E^{-1}}_+^{\hat{\mu}\hat{\nu}}\partial_{\hat{\mu}}\partial_{\hat{\nu}}
		X^{\hat{M}} +\frac{1}{\sqrt{-\det
		E_{\hat{\rho}\hat{\sigma}}}}\partial_{\hat{\mu}}\corchet{\sqrt{-\det
		E_{\hat{\rho}\hat{\sigma}}}\parent{E^{-1}}_+^{\hat{\mu}\hat{\nu}}}
		\partial_{\hat{\nu}} X^{\hat{M}} \nonumber\\ && \hspace{10mm}
		+\parent{E^{-1}}_+^{\hat{\mu}\hat{\nu}}\partial_{\hat{\mu}}X^{\hat{K}}
		\partial_{\hat{\nu}}X^{\hat{L}}\Gamma_{\hat{K}\hat{L}}{}^{\hat{M}}=0,
		\label{geneq1} \\ 
&& 	\partial_{\hat{\mu}}\corchet{\sqrt{-\det
		E_{\hat{\rho}\hat{\sigma}}}\parent{E^{-1}}_-^{\hat{\mu}\hat{\nu}}} =0
		\label{geneq2}
\end{eqnarray}
where $\parent{E^{-1}}_+$ and $\parent{E^{-1}}_-$ are the symmetric and
antisymmetric parts of $E^{-1}$ respectively and
$\Gamma^{\hat{M}}_{\hat{K}\hat{L}}$ are the Christoffel symbols of the
metric $g$ on the target space. The first equation is obtained by taking
variations of \ec{covactionA} with respect to $X^{\hat{M}}$ and the second
one by taking variations with respect to $A_{\hat{\mu}}$.

We assume the foliation of the world-volume $\Sigma={\cal
M}\times\mathbb{R}$, where ${\cal M}$ is a Riemannian manifold. We
consider ${\cal M}$ to be a compact manifold with no boundaries. We impose
the following conditions
\begin{eqnarray}
&&  X^0=\sigma^0, \hspace{10mm} X^M=X^M\parent{\sigma^\alpha}, \label{static1} \\
&&  g_{00}=-1, \hspace{10mm} g_{0M}=0, \label{static3} \\
&&  A_0=0,\hspace{13mm} A_\mu=A_\mu\parent{\sigma^\alpha}, \label{static2}
\end{eqnarray} such that the remaining fields $X^M$ and
$A_\mu$ are static. The resulting equations over ${\cal M}$ are
\begin{eqnarray}
&&  \parent{E^{-1}}_+^{\mu\nu} \partial_\mu\partial_\nu X^M +\frac{1}{\sqrt{\det
		E_{\rho\sigma}}}\partial_\mu\corchet{\sqrt{\det
		E_{\rho\sigma}}\parent{E^{-1}}_+^{\mu\nu}}\partial_\nu X^M \nonumber \\
&&	\hspace{10mm} +\parent{E^{-1}}_+^{\mu\nu}\partial_\mu X^K \partial_\nu X^L
		\Gamma_{KL}{}^M=0, \label{eq1} \\ 
&& \partial_{\mu}\corchet{ \sqrt{\det
		E_{\rho\sigma}}\parent{E^{-1}}_-^{\mu\nu}} =0. \label{eq2}
\end{eqnarray}
We also suppose that some sector of the target space is compactified, that
is the spatial part of the target space is the product of a compact
Riemannian manifold $T$ times a hyperplane. We assume $\dim T \geq \dim
{\cal M}$. We fix those components of the map corresponding to the
hyperplane equal to zero. Therefore, the remaining degrees of freedom are
$X^r$, $r=1,\ldots,\dim T$, and $A_\mu$, being both fields over ${\cal
M}$. We consider $X^r$ to be the components of an immersion ${\cal
M}\rightarrow T$, therefore $G_{\mu\nu}=\partial_\mu X^r
\partial_\nu X^sg_{rs}$ is a Riemannian metric on ${\cal M}$.

Now we introduce the K\"ahler geometry into the problem. Let ${\cal M}$ be
a K\"ahler manifold of real dimension $p=2n$. Let $J$, $\Omega$ and
$K\parent{U,V}=\Omega\parent{U,JV}$ be the almost complex structure, the
K\"ahler form and the K\"ahler metric of ${\cal M}$ respectively. These
three objects are regular tensors over ${\cal M}$, the associated matrices
in a local coordinate system being invertible. By definition $\Omega$ is
closed and implicitly co-closed with respect to the K\"ahler metric,
hence it is harmonic with respect to this metric. We suppose that
$A_{\mu}$, $X^r$ and $g_{rs}$ verify \barr \label{ansatzG} &&
G_{\mu\nu}=K_{\mu\nu} ,
\\ && F_{\mu\nu}=\Omega_{\mu\nu}.\label{ansatzF}\earr If
\ec{ansatzG} and \ec{ansatzF} are verified, the Born-Infeld tensor becomes
$ E_{\mu\nu}=K_{\mu\nu}+\Omega_{\mu\nu}$ and then we get \begin{equation}
\label{deter} \det E_{\mu\nu}=2^n\det K_{\mu\nu}
,\hspace{8mm}\parent{E^{-1}}_+^{\mu\nu}=\frac{1}{2}K^{\mu\nu}
,\hspace{8mm}\parent{E^{-1}}_-^{\mu\nu} = \frac{1}{2}\Omega^{\mu\nu}\end{equation}
where $K^{\mu\nu}$ and $\Omega^{\mu\nu}$ are the inverses of $K_{\mu\nu}$
and $\Omega_{\mu\nu}$ respectively, also
$\Omega^{\mu\nu}=-K^{\mu\alpha}K^{\nu\beta}\Omega_{\alpha\beta}$.

Note that eqs. \ec{eq2} are automatically solved by \ec{ansatzG} and
\ec{ansatzF}. Indeed, if we substitute \ec{deter} in \ec{eq2}, the LHS of the resulting expression can be brought to $d^\star\Omega$, where $^\star$ is the Hodge dual
constructed with $K$, and this expression is identically zero. Equations
\ec{eq2} represent the time-independent Born-Infeld equations and the fact
that the configuration \ec{ansatzG}-\ec{ansatzF} solves them is a
particular case of the more general result about extended self-dual
configurations, as we have already mentioned.

We may concrete the solution by choosing an explicit gauge field $A$ such
that its curvature is the K\"ahler form of ${\cal M}$. We remark that for
compact manifolds the K\"ahler form is closed and not globally exact.
Therefore the gauge field must be given by local configurations with non-trivial transitions, as expected for the case of a gauge field derived
from a connection on a non-trivial principle bundle. 

The K\"ahler form of any K\"ahler manifold may be locally expressed in terms of the K\"ahler
potential ${\cal K}$, 
\begin{equation}\label{KahlerPotential}
  \Omega={\rm i}\partial_a\partial_{\bar{a}}{\cal K} dz^a\wedge
  d\bar{z}^{\bar{a}}
\end{equation}
where we are using complex coordinates. By locality we mean that ${\cal K}$ is
not a scalar function over the whole manifold ${\cal M}$. For a given K\"ahler
manifold one has a collection of K\"ahler potentials, one for each chart
with a generic expression in terms of the coordinates. In the overlapping
of two charts the coordinates are related by a holomorphic coordinate
transformation, under which the K\"ahler potential transforms
inhomogeneously with holomorphic and antiholomorphic functions \begin{equation}
{\cal K}'(z'(z),\bar z'(\bar z)) = {\cal K}\parent{z,\bar z}+\phi(z)+\psi(\bar z).
\end{equation} This transformation does not affect the global character of the
K\"ahler form given in \ec{KahlerPotential}. 

The K\"ahler potential provides an expression for the gauge field. In fact, on each chart we may introduce the one-form
\begin{equation}
  A=\frac{{\rm i}}{2}\parent{\partial_{\bar{a}} {\cal K} d\bar{z}^{\bar{a}} -
  \partial_a {\cal K} dz^a} \label{gaugefield}
\end{equation}
which verifies $\Omega = dA$. Under a holomorphic coordinate
transformation we have \begin{equation} A' = A + d \corchet{\frac{{\rm i}}{2}
\parent{\psi-\phi}}\label{gauge} \end{equation} which is a $U(1)$ gauge
transformation. Eq. \ec{gauge} ensures that the collection of fields given
in \ec{gaugefield} are the local pullbacks of a regular connection on a
non trivial $U(1)$ principle bundle over ${\cal M}$, the pullbacks are
constructed with a set of local sections.

Once the analysis of the gauge field has been completed, we may study the
remaining fields $X^r$ in eqs. \ec{eq1} and \ec{ansatzG}. Under
\ec{ansatzG}-\ec{ansatzF}, equations \ec{eq1} yields
\begin{equation}\label{fineq}
  K^{\mu\nu}\partial_\mu\partial_\nu X^r -K^{\mu\nu} \partial_\sigma X^r\gamma^\sigma_{\mu\nu}
  +K^{\mu\nu}\partial_\mu X^t \partial_\nu X^u\Gamma_{tu}^r=0
\end{equation}
where $\gamma$ is the Christoffel symbol of the K\"ahler metric $K$. The
first two terms describe the laplacian, defined with the K\"ahler metric,
applied to $X^r$. This equations represent the classical problem of
harmonic mapping between $({\cal M},K_{\mu\nu})$ and $(T,g_{rs})$.
Moreover, considering that the K\"ahler metric must be induced by the maps
as it is stated by \ec{ansatzG}, the minimal map $X^r$ actually represents
a minimal immersion of the K\"ahler manifold ${\cal M}$ into $T$, that
is, it is a minimum of the volume action
\[ \int\limits_{{\cal M}} d^p\sigma\sqrt{\det K_{\rho\sigma}} \]
where
\begin{equation}
   K_{\rho\sigma} = \partial_\rho X^r \partial_\sigma X^s g_{rs}.
\end{equation}
Therefore we have characterized a class of solutions of D-branes by
setting the world volume of the D-brane to be a K\"ahler manifold, the
gauge field being a connection whose curvature is the K\"ahler form and
the scalar fields realizing a minimal immersion of the K\"ahler manifold
into the target space. Moreover it is straightforward to show that
\ec{eq1} and \ec{eq2} are exactly the same equations to be satisfied by
the minima of the Hamiltonian of the Dirac-Born-Infeld action in the case
of a flat target space.

So far we were focused in the K\"ahler geometry of the base manifold
${\cal M}$, but the topology and geometry of the compactified target space $T$ was
left unspecified. Now we may study specifical configurations for $T$. In
this work we take the case in which $T$ is a flat torus, which in term
are K\"ahler manifolds. A lot of the work done in String Theory and M
theory with compactified target consider the cases of Calabi-Yau
manifolds, being the tori the simplest, non trivial examples among them.
Also orbifolds which are the quotient space of tori by discrete symmetries have been extensively considered. We hope to extend, in future work, our quantization arguments of next sections to include also these target spaces.  

In general the existence of minimal immersions of K\"ahler manifolds into tori is a non trivial subject and some particular constructions have been studied in mathematical works imposing several constraints on the topology of the base K\"ahler manifold. The first non-trivial example at hand is when the base manifold ${\cal M}$ is a Riemmanian surface of genus $g$ and the torus is a flat one of dimension $2g$, in this case a minimal immersion is given in terms of holomorphic one-forms over ${\cal M}$. We study this example in detail in the next section. The extension of this minimal immersion for higher dimensional K\"ahler base manifolds is the Albanese map of a compact K\"ahler manifold whose cotangent bundle is ample \cite{Matsushima,Nagano}, this condition means that the holomorphic one-forms on ${\cal M}$ determine the full cotangent space at each point of ${\cal M}$. The Albanese map of a compact K\"ahler manifold
is an holomorphic map and the associated Albanese torus is a flat
torus whose dimension is just the number of independent holomorphic
one-forms on the base manifold, $h^{1,0}({\cal M})$. In particular any
compact Riemannian surface of positive genus $g$ satisfies the condition of ample cotangent bundle.

\section{The $D=11$ Supermembrane wrapped on a calibrated submanifold}
\label{Membrane} 
\subsection{Topological conditions on the configuration space}\label{PhaseSpace}
The $D=11$ toroidal Supermembrane wrapped on the compact sector of a 
$M_9\times S^1 \times S^1$ target space, where the configuration space 
satisfies a topological condition implying an \emph{irreducible} wrapping 
\cite{IM+AR+RT} or equivalently a non-trivial central charge on the 
supersymmetric algebra was analyzed in \cite{LB+MG+IM+AR,LB+MG+AR,IM+AR+RT,IM+JO+AR}. 
It was proven in \cite{LB+MG+IM+AR,LB+MG+AR} that the spectrum of its regularized 
Hamiltonian is discrete, in \cite{LB+AR} the construction of its heat kernel and the 
associated Feynmann formulas were presented. Those constructions were based on the 
property of the supersymmetric Hamiltonian, under the topological restrictions, 
of being compatible with its bosonic sector, in the precise sense given in \cite{LB+AR}.

We would like to generalize the above constructions in order to consider more general compactified sectors of the target space. It is crucial in obtaining a discrete spectrum
to have the topological condition which is equivalent
to have a a non-trivial flux on the base manifold. As we will see, this fact is related
to the existence of a minimal immersion of the base manifold in the target space. 
The minimal immersion is of course a minimum of the Hamiltonian. It turns out that the image of the Supermembrane is a calibrated submanifold of the compactified sector of the target space. The minimal immersion becomes straightforward in
the case of a toroidal Supermembrane wrapped on $M_9\times S^1\times S^1$ and
therefore it is implicit in \cite{LB+MG+IM+AR,LB+MG+AR,IM+AR+RT,IM+JO+AR}. 

Similarly to the previous section, we consider the supermembrane in a target space which factorizes as a Minkowski manifold times a compact Riemannian one, this target being
a solution of pure $D=11$ Supergravity. The compact sector of the target space is
a K\"ahler manifold admitting a holomorphic inmersion of the base manifold.

We consider the $D=11$ Supermembrane in the light cone gauge \cite{deWitt}, 
where the light cone coordinates $X^+$ and $X^-$ involve $X^0$ and one of non-compactified directions, 
\begin{eqnarray}
&&	X^+ = P_0^+\tau \\
&&	P^+ = P_0^+\sqrt{w} \\
&&	\Gamma^+ \theta =0
\end{eqnarray}
where $\sqrt{w}$ is a scalar density independent of $\tau$. 
In general it is not possible to take $\sqrt{w}=1$ on the whole compact base manifold. 
This density is preserved by the residual gauge symmetry of the theory and determines 
the element of area of the base manifold. Later on we will 
relate this density to the metric induced by a minimal configuration.

The dynamics of the Supermembrane in the light cone gauge is described by the scalar fields $X^M$, $M=1,\ldots,9$, over the base manifold ${\cal M}$ and their canonical conjugate variables $P_M$. $X^M$ are maps from a compact Riemann surface ${\cal M}$ of genus $g>0$ to the target space $M\times T$, where $M$ is a Euclidean subspace and $T$ is a flat torus. 
The explicit holomorphic immersion we are going to use 
requires the flat torus to have dimension $2g$.

The bosonic Hamiltonian of the Supermembrane in the light cone gauge \cite{deWitt} is given by
\begin{equation}\label{smham}
	H = \int\limits_{\cal M}
	d^2\sigma\sqrt{w} \corchet{\frac{1}{2w}P_M^2
	+ \frac{1}{4}\{X^M,X^N\}^2}\;,
\end{equation}
subject to the first class constraint
\begin{equation}\label{GenConstraint} 
	\oint\limits_{\cal C} \frac{1}{\sqrt{w}} P_M d X^M = 0
\end{equation}
for any closed path ${\cal C}$ on ${\cal M}$. This constraint generates the area 
preserving diffeomorphisms, APD, over ${\cal M}$. The brackets in (\ref{smham}) are the symplectic ones defined in (\ref{bracket}) using the symplectic structure
\begin{equation}
	\frac{\epsilon^{\mu\nu}}{\sqrt{w}}\;.
\end{equation}

The first class constraint is equivalent to the following two statements: \\
i) The integrand in
\ec{GenConstraint} is a closed one form, that is 
\begin{equation} \label{InfiAPD}
  \phi\equiv\{P_M, X^M\}=0
\end{equation}
ii)
\begin{equation}\label{GlobalAPD}
  \hat\phi \equiv \oint\limits_{{\cal C}_r} \frac{1}{\sqrt{w}} P_M d X^M = 0\;, 
\end{equation}
where ${\cal C}_r$, $r=1,\ldots,2g$, is a basis of homology on ${\cal M}$. 
$\phi$ generates APD connected to the identity while $\hat\phi$ generates APD non-connected to the
identity. 

We may now introduce the configuration space of the $D=11$ Supermembrane with 
non-trivial central charge arising from the wrapping on a two cycle. 
We consider the flat torus $T$ to be $S^1\times\cdots\times S^1$ ($2g$ times). The maps to each $S^1$ are given by
\begin{equation}
  P\in {\cal M} \rightarrow \int\limits_{P_0}^P dX^r \; {\rm modulo\;integers}
\end{equation}
where $dX^r$ satisfy
\begin{eqnarray}
&&	\oint\limits_{{\cal C}_s} dX^r={\rm integers} \label{torus} \\
&&	\int\limits_{\cal M} dX^r\wedge dX^s =n\omega^{rs}\;, \hspace{10mm} n\neq 0 \label{winding}
\end{eqnarray}
where $P_0$ is a fixed point in ${\cal M}$, $n$ is a non-zero integer, and 
$\omega^{rs}$ is the canonical symplectic matrix on $T_g$, associated to the K\"ahler form on $T_g$.

Let us discuss several interpretations of the above condition, which plays a central role in our analysis. Eq. (\ref{winding}) is a ``topological" condition in the sense that it restricts 
the configuration space to a definite principle bundle. In fact,
$\frac{1}{2} \omega_{rs} dX^r\wedge dX^s$ is a closed two-form whose integral over
 ${\cal M}$ is a integer $n$, hence there exits a non-trivial $U(1)$ principle 
bundle over ${\cal M}$ and a connection over it such that the above two-form 
is its curvature. Moreover, the principle bundles over the compact Riemann surface 
${\cal M}$ are classified by the integer $n$, by considering the configurations of 
the Supermembrane to satisfy (\ref{winding}) we are then quantizing the Supermembrane 
in a given principle bundle, characterized by $n$. This two-form may also be 
interpreted as a non-trivial flux condition on the dynamics of the Supermembrane.

Let us recall that the SUSY algebra for the $D=11$ Supermmebrane 
with winding exhibits central charges \cite{deWitt+Peeters+Plefka},
\begin{equation}
  \parent{ Q^+_\alpha, \bar
  Q^-_\beta}_{DB} = - \parent{ \gamma_M \gamma_+ \gamma_-}_{\alpha\beta} P_0^M -
  \frac{1}{2} \left(\gamma_{rs} \gamma_+ \gamma_-\right)_{\alpha\beta} \int\limits_{\cal M}
  d^2\sigma\sqrt{w}\{X^r, X^s\}
\end{equation}
and we notice that the last factor of the right hand member is 
equal to the left hand member of (\ref{winding}). 

Eq. (\ref{winding}) may also be interpreted in the following way: Let us first consider $g=1$, that is $r=1,2$. Then the condition (\ref{winding}) ensures that the image by $X^r$ is a torus 
which does not degenerate to a circle, for this reason (\ref{winding}) was called in \cite{IM+AR+RT} the condition of irreducible wrapping. For $g\geq 1$ the interpretation is the  following: The minimal immersions of ${\cal M}$ into $T$ we are going to introduce in the next subsection, and whose existency in general is a nontrivial subject, define calibrated submanifolds in $T$, then the topological condition (\ref{winding}) states that the supermembrane wraps on the calibrated submanifold. Calibrated submanifold and intersecting branes are related subjects; for $g>1$ condition (\ref{winding}) implies that the ground state of the system may be interpreted as an intersecting brane solution, where non-degenerate toroidal branes intersect at the time direction. We will analyze this point in subsection 3.3. 

The configuration space of the Supermembrane is completed by taking 
$X^m$, $m=2g+1,\ldots,9$, to be singled valued over ${\cal M}$ and hence satisfying 
\begin{eqnarray}
&&		\oint\limits_{{\cal C}_r} dX^m=0\;, \\
&&		\int\limits_{\cal M} dX^m\wedge dX^r =0\;,\hspace{10mm} 
		\int\limits_{\cal M} dX^m\wedge dX^n =0\;.
\end{eqnarray}
They may then be interpreted as maps from the base manifold to the topologically trivial
Euclidean subspace $M$ of the target space.

We notice that condition (\ref{winding}) does not exclude string configurations on 
the space of configurations defined above. In fact, consider a particular $dX^r$ 
satisfying (\ref{winding}) and $X^m$ depending only on a linear combination of 
$\sigma^1$ and $\sigma^2$. It is not difficult to see, for example for the $g=1$ case,
that those configurations correspond to a ten dimensional string wrapped on $S^1$ \cite{NH+AR+JS}. These configurations contribute with a finite energy to the Hamiltonian. There are no string-like spikes in contrast with the Supermembrane without the topological restrictions \cite{deWitt+Luscher+Nicolai,deWit+M+N,deWitt+Peeters+Plefka}. 
The string-like spikes contribute with zero energy to the Hamiltonian and 
it is not possible to know if they are attached or not to the Supermembrane. It is for this reason that the Supermemebrane without the topological restrictions must be interpreted as a many extended objects theory.

\subsection{Harmonic forms, minima of the Hamiltonian and calibrated submanifolds}\label{Immersion}
The harmonic one-forms play a relevant role in our formulation because they 
are minima of the Hamiltonian and also all the topological information of 
the configuration space is contained in them, as we will see.

The local isolated minima of the Hamiltonian (\ref{smham}) satisfy
\begin{equation} 
	P_m =P_r=0, \hspace{10mm} X^m=\mathrm{constant} 
\end{equation} 
and 
\begin{equation}\label{MinimalImmersion} 
	G^{\nu\mu}\partial_\mu\partial_\nu X^r -G^{\mu\nu}
	\partial_\rho X^r\gamma_{\mu\nu}{}^\rho =0 
\end{equation}
where, as in the section
\ref{Solutions}, $G_{\mu\nu}$ is the induced metric and
$\gamma_{\mu\nu}{}^\rho$ its Christoffel symbols. Last equations are the
same equations \ec{fineq}, with no contribution coming from the curvature
of the target because it is flat. Therefore, the minima of the Hamiltonian
\ec{smham} represent minimal immersions of the Supermembrane into the
target. Moreover any Riemannian manifold is K\"ahler, so this problem is
equivalent to the one we consider in section \ref{Solutions}. We remark
that eq. \ec{MinimalImmersion} is equivalent to demand \begin{equation} d\:^\star dX^r=0\end{equation}
 and this implies that the closed one-forms $dX^r$ associated to each solution of
\ec{MinimalImmersion} are harmonic over ${\cal M}$. 
This is not a general result for harmonic maps between
Riemannian manifolds, it holds for the particular case we are considering
when the target is flat. The minimal maps we consider are then harmonic
maps satisfying the restriction (\ref{winding}).

We now proceed to construct the minimal map we are interested in. We denote by
$\left(\alpha_i,\beta_i\right)$, $i=1,\ldots,g$, a canonical basis of homology over
${\cal M}$ and by $\omega_i$ a basis of holomorphic
one-forms over ${\cal M}$ normalized by
\begin{equation}
   \oint\limits_{\alpha_j} \omega_i = \delta_{ij}\;,\;\;\;\;
   \oint\limits_{\beta_j} \omega_i = \Pi_{ij}
\end{equation}
where $\Pi$ is the period matrix, it is symmetric with positive definite
imaginary part $\Im{\rm m}\Pi$. We rewrite
\begin{equation}
  \omega_i=d\tilde X_i^1+{\rm i}d\tilde X_i^2\;,
\end{equation}
where $d\tilde X_i^I$, $I=1,2$, are harmonic one forms over ${\cal M}$. In constructing the 
minimal map, we decompose the compact target index as $r=(i,I)$. Notice that $\tilde X^I_i$ is 
not singled-valued over ${\cal M}$.

We introduce the following set of harmonic one-forms
\begin{equation}
  d\hat X_i^I \equiv \left(M^{-1}d\tilde X\right)_i^I
\end{equation}
where $M$ is, as a matrix in the $I,J$ indices
\begin{equation}
  M_{ij}=\left(\begin{array}{cc} \delta_{ij} & \Re{\rm e} \Pi_{ij} \\
                            0 & \Im{\rm m} \Pi_{ij}\end{array}\right)
\end{equation}
and
\begin{equation}
  \left(M^{-1}\right)_{ij}=
    \left(\begin{array}{cc}
          \delta_{ij} & -\Re{\rm e} \Pi_{ik} \left(\Im{\rm m}\Pi\right)^{-1}_{kj} \\
          0 & \left(\Im{\rm m} \Pi\right)^{-1}_{ij}\end{array}\right)\;.
\end{equation}
We get
\begin{equation}
  \oint\limits_{{\cal C}_j^J} d\hat X_i^I = \delta^{IJ}\delta_{ij}
\end{equation}
where ${\cal C}_j^1=\alpha_j$ and ${\cal C}_j^2=\beta_j$.

The minimal immersion from ${\cal M}$ to $T$ is defined by
\begin{equation}\label{MinimalMap} 
  P\in {\cal M}\rightarrow \int\limits_{P_0}^P d\hat X_i^I\;\;{\rm modulo\;integers}\;. 
\end{equation}
This map verifies the topological condition (\ref{winding}), with $n=1$.

We now introduce the metric on the flat torus by
\begin{equation}
  ds^2 = 2 \left(\Im{\rm m}\Pi\right)^{-1}_{ij} dZ_i\otimes d\bar Z_j\;,
\end{equation}
the factor $\left(\Im{\rm m}\Pi\right)^{-1}$ is necessary when considering the induced 
metric on ${\cal M}$ from the target metric, in order to
obtain invariance under canonical changes of the canonical 
homology basis on ${\cal M}$. We obtain for the induced metric
\begin{eqnarray}
   ds^2 &=& 2 \left(\Im{\rm m}\Pi\right)^{-1}_{ij} \omega_i\otimes \bar
        \omega_j \\
        &=& 2 \left(\Im{\rm m}\Pi\right)^{-1}_{ij} \delta^{IJ} d\tilde
        X_i^I \otimes d\tilde X_j^J \\
        &=& 2g_{ij}^{IJ} d\hat X_i^I\otimes d\hat X_j^J
\end{eqnarray}
where
\begin{equation}
  g\equiv M^T \left(\begin{array}{cc}
     \left(\Im{\rm m}\Pi\right)^{-1} & 0 \\
     0 & \left(\Im{\rm m}\Pi\right)^{-1} \end{array}\right) M
\end{equation}
is symmetric, positive definite and with $\det g=1$. 
Although the immersion depends on the canonical basis of homology 
$(\alpha,\beta)$, the induced metric is invariant under changes of canonical basis.

Let us now define $\sqrt{w}$ 
\begin{eqnarray}
  \sqrt{w} d\sigma^1\wedge d\sigma^2 &\equiv& 
  \frac{1}{2}\epsilon_{IJ} d\hat X^I_i \wedge d\hat X^J_j \delta^{ij} \\  
  &=& \frac{\partial Z_k}{\partial z} \frac{\partial \bar Z_l}
  {\partial \bar z} \left(\Im{\rm m}\Pi\right)^{-1}_{kl} \frac{1}{2{\rm i}} dz\wedge d\bar z
\end{eqnarray}
where $z=\sigma^1+ {\rm i}\sigma^2$ is the local complex coordinate over ${\cal M}$ and 
\begin{equation}
	Z_i(z)= \int\limits_{z_0}^z \omega_i\;.
\end{equation}
It is clear that $\sqrt{w}$ is equal to the square root of the induced K\"ahler metric.

The image of the Supermembrane in the flat torus throughout the minimal map is a calibrated submanifold of the torus. To see this, we first remark that the minimal, holomorphic map induces on the base manifold both the K\"ahler form and the K\"ahler metric form the target space, as it is indicated in the appendix. The following two form 
\begin{equation}
	\omega_{rs} d\hat X^r \wedge d\hat X^s\;,
\end{equation}
which is just the induced K\"ahler form over ${\cal M}$, represents a calibration on the image of ${\cal M}$ in $T$ since it is equal to the induced volume element at each point of ${\cal M}$.

\subsection{Unbroken Supersymmetries}
In this section we analyze the remainning supersymmetry after introducing 
the minimal configuration. We study the cases $g=1$ and $g=2$.

The minimal configuration has bosonic sector given in (\ref{MinimalMap}) and vanishing fermionic sector. In 
order the minimal configuration to be a supersymmetric configuration it is neccessary that
\begin{equation}
	\delta\theta=0\;,
\end{equation}
where $\theta$ is the spinor field over the world-volume.

In order to determine the number of supersymmetries left unbroken it 
is convenient to use the following kappa symmetry gauge fixing condition
\begin{equation}\label{Kappafixing}
	\left(1+\Gamma\right)\theta = 0
\end{equation}
where
\begin{eqnarray}
&&	\Pi_{\hat\mu}{}^{\hat K} = \partial_{\hat\mu} X^{\hat K}  -
		{\rm i} \bar\theta\Gamma^{\hat K}\partial_{\hat\mu}\theta\;, \\
&&	\Gamma = \frac{i}{6\sqrt{-G}} \epsilon^{\hat\mu\hat\nu\hat\rho} 
		\Pi_{\hat\mu}{}^{\hat K} \Pi_{\hat\nu}{}^{\hat L} \Pi_{\hat\rho}{}^{\hat M}
		\Gamma_{\hat K\hat L\hat M} \\
&&	\Gamma^2=1
\end{eqnarray}
Condition (\ref{Kappafixing}) is preserved only if supersymmetric 
transformations are compensated
with a kappa transformation, hence we have the combined supersymmetry and kappa symmetry
\begin{equation}
	\delta\theta = \left(1+\Gamma\right)\kappa +\varepsilon=0
\end{equation}
where $\kappa$ is a constant spinor satisfying
\begin{equation}
	\left(1+\Gamma\right)\kappa = -\frac{1}{2}\left(1+\Gamma\right)\varepsilon\;.
\end{equation}
We then have the equation defining a supersymmetric configuration
\begin{equation}\label{SUSYCondition}
	\left(1-\Gamma\right)\varepsilon =0\;.
\end{equation}

We may compute explicitly the matrix $\Gamma$ for the minimal configuration
\begin{equation}\label{Gammaepsilon}
	\Gamma\varepsilon =-\frac{{\rm i}}{2\sqrt{-G}} P_0^+ \partial_{\mu} \hat X^r 
	\partial_{\nu} \hat X^s \epsilon^{\mu\nu} \Gamma_{0rs}\varepsilon\:.
\end{equation}

In the $g=1$ case the equation (\ref{Gammaepsilon}) takes the form
\begin{equation}
	\Gamma\varepsilon = -{\rm i}\Gamma_{012}\varepsilon\;,
\end{equation}
hence the solution to (\ref{SUSYCondition}) is
\begin{equation}
	\Gamma_{012}\varepsilon = {\rm i}\varepsilon\;,
\end{equation}
in the latest we have used tanget space indices. This condition preserves one half of the 
original supersymmetries.

In the $g=2$ case, a solution to (\ref{SUSYCondition}) is given by the global conditions
\begin{eqnarray}
&&	\Gamma_{013}\varepsilon = {\rm i}\varepsilon \label{SUSYCond1}\\
&&	\Gamma_{024}\varepsilon = {\rm i}\varepsilon \label{SUSYCond2} \:.
\end{eqnarray} These conditions are
equivalent to substitute the expression $\Gamma_{0rs}\varepsilon$ by 
${\rm i} \omega_{rs}\varepsilon$ in (\ref{Gammaepsilon}) and then we end up with a solution to (\ref{SUSYCondition}). To see how this works out it is neccessary to use 
the holomorphic properties of the minimal immersion. Conditions (\ref{SUSYCond1}) and 
(\ref{SUSYCond2}) are the usual ones for two toroidal $M2$-branes intersecting at the time 
direction. This configuration preserves one quarter of the original supersymmetry.

\subsection{Symplectic Yang-Mills Theory from wrapped $D=11$ Supermembranes}
\label{SymplecticYangMills}
Let us now consider the most general map between ${\cal M}$ and $T$. It may be written 
in terms of the above basis $d\hat X^I_i$ of harmonic one-forms. Any closed one-form 
may be expressed as 
\begin{equation}\label{prechange}
	dX^I_i = S^{IJ}_{ij}d\hat{X}^J_j+{\rm exact\;one\textrm{-}form.}
\end{equation} 

In order that $dX^I_i$ satisfy restriction (\ref{torus}) we must have
\begin{equation}
	\int\limits_{{\cal C}^J_j} dX^I_i = S^{IK}_{ik} \delta^{KJ}\delta_{kj} = S^{IJ}_{ij}\;,
\end{equation}
therefore the coefficients $S^{IJ}_{ij}$ have to be integers. 

In order that $dX^I_i$ satisfy (\ref{winding}) we must have
\begin{equation}
	\int\limits_{\cal M} dX^I_i \wedge dX^J_j = S^{IK}_{ik} 
	\epsilon^{KL} \delta_{kl} S^{JL}_{jl} = \epsilon^{IJ}\delta_{ij}\;.
\end{equation}
Consequently, $S^T$ and $S$ belong to $Sp(2g,\mathbb{Z})$. Moreover
$Sd\hat X$ is the canonical basis of harmonic one-forms associated to the homology basis 
$\left(S^T\right)^{-1} {\cal C}$. In fact $\left(S^T\right)^{-1},S^{-1}\in 
Sp(2g,\mathbb{Z})$ and 
\begin{equation}
	\int\limits_{\left( S^T\right)^{-1\,JK}_{jk} {\cal C}^K_k} 
	S^{IL}_{il} d\hat X^L_l =	\delta^{IJ} \delta_{ij}\;.
\end{equation} 
Then the general map (\ref{prechange}) is constructed with the matrix $S$ 
corresponding to a change on the harmonic basis associated to the diffeomorphisms 
mapping a canonical homology basis into another one. Under those maps and 
using the previous expression for $\sqrt{w}$ we obtain
\begin{eqnarray}
	\sqrt{w'(\sigma)} &=& \frac{1}{2} \epsilon_{IJ} 
	\frac{\partial \hat X'^I_i} 			
	{\partial\sigma^a} \frac{\partial \hat X'^J_j} 
	{\partial\sigma^b} \delta^{ij}\epsilon^{ab} \\
	&=& \left(S^T\epsilon \delta S\right)^{KL}_{ij} \frac{1}{2} \frac{\partial \hat X^I_i} 				{\partial\sigma^a} \frac{\partial \hat X^J_j} {\partial\sigma^b} \epsilon^{ab} \\
	&=& \frac{1}{2}\epsilon_{KL}\delta^{kl} \frac{\partial \hat X^K_k}{\partial\sigma^a} \frac{\partial \hat X^L_l} {\partial\sigma^b} \epsilon^{ab} = 
	\sqrt{w(\sigma)}\;.
\end{eqnarray}
The diffeomorphisms corresponding to $S$ are then area preserving and non connected to the identity, and consequently $S$ may be fixed by using this kind of APD which are part of the residual 
gauge invariance of the theory in the light cone gauge and whose generator is (\ref{GlobalAPD}). We may then take, after 
rescaling of the exact one-forms $d{\cal A}^I_i$ by the constant metric 
$g^{IJ}_{ij}$ and returning to the single compact target index $r$,
\begin{equation}\label{change} 
	dX^r = d\hat X^r+d{\cal A}^r
\end{equation}
The physical degrees of freedom are only contained in ${\cal A}_r\equiv g_{rs} {\cal A}^s$. 
However, the variables ${\cal A}_r$ still have the gauge invariance asociated to the APD connected to the identity whose generator is (\ref{InfiAPD}). As we are going to see, this gauge symmetry correspond to a symplectic gauge symmetry and therefore we take ${\cal A}_r$ as a symplectic gauge connection in the way explained in the appendix.

We now consider \ec{change} in the dynamics of the Supermembrane. We first
note that the kinetic terms $P_M\dot{X}^M$ of the canonical action may be
rewritten as \begin{equation} P_r\dot{X}^r + P_m\dot{X}^m = P^r\dot{{\cal
A}}_r+P_m\dot{X}^m \end{equation} since $\hat X^r$ do not depend on the world-volume
time. Therefore $P^r$ are the conjugate momenta to ${\cal A}_r$.
Furthermore, the APD constraint \ec{InfiAPD} acquires the following form
\begin{equation}\label{Gauss} 
	\phi = {\cal D}_r P^r +\llaves{X^m,P_m} 
\end{equation} 
and is now interpreted as the Gauss constraint generating the symplectic gauge
symmetry. 

The final form of the bosonic Hamiltonian is
\begin{eqnarray}\label{NCYM}
  H &=& \frac{1}{2}\int\limits_{\cal M} d^2\sigma\sqrt{w}  \left[\frac{1}{w}\parent{P_m^2+
  {P^r}^2} + \frac{1}{2}{\cal F}_{rs}^2+ g^{rs}\delta_{mn}{\cal D}_rX^m {\cal D}_sX^n \right. \nonumber \\
  & &
  \hspace{50mm} \left. +\frac{1}{2}\{X^m,X^n\}^2 + n^2\right]
\end{eqnarray}
where we have imposed the global condition
\begin{equation}\label{trivial}
	\int\limits_{{\cal M}} {\cal F}_{rs} d\hat X^r \wedge d\hat X^s
	=0\;.
\end{equation}
The Hamiltonian (\ref{NCYM}) correspond to a symplectic 
Yang-Mills Theory on the image of the base manifold in the 
compact sector of the target space, which is a calibrated submanifold. 
The symplectic Yang-Mills field is 
coupled to the transverse scalar
fields $X^m$. The condition 
(\ref{trivial}) is analogous to the condition in standard 
abelian Yang-Mills of considering 
connections on a particular principle bundle. 

\subsection{Discreteness of the spectrum of the regularized Hamiltonian}
The degrees of freedom of the above Hamiltonian are the maps $X^m$ and 
${\cal A}_r$ and their conjugate momenta. All these variables 
are singled-valued over ${\cal M}$, we have shown that due to the 
topological condition (\ref{winding}) there are not degrees 
of freedom associated to the harmonic sector. We now proceed to 
consider a regularization of the above Hamiltonian. Following 
\cite{deWit+M+N,Hoppe} we introduce a basis $Y_A(\sigma)$ on the space of scalar functions 
on the base manifold and we decompose all the variables of the phase space as
\begin{eqnarray}
&&	\phi(\sigma) = \sum\limits_{A} \phi^A Y_A(\sigma)\;, \\
&&	\{Y_A(\sigma),Y_B(\sigma)\} = f_{AB}{}^C Y_C(\sigma)	
\end{eqnarray}
where the above bracket is the symplectic one. $f_{AB}{}^C$ are the 
structure constants of the APD of the 
base manifold. These structure constants correspond to the 
ones of $su(N)$ in the limit $N\rightarrow\infty$. 
Then we take a truncation of the theory 
by considering $N$ finite. After regularization the Hamiltonian 
becomes a quantum mechanical Schr\"oedinger operator defined on a real space. 
All the topological information of the base manifold is 
contained in the basis $Y_A(\sigma)$ of the space of 
functions over the base manifold.

The potential of the regularized Hamiltonian satisfies the following 
two properties: \\
i) $V(X,{\cal A}) = 0$ if and only if $X^m_A=0$ and ${\cal A}_{rA}=0$ \\
ii) $V(X,{\cal A})\rightarrow\infty$  for $X^m_A,{\cal A}_{rA} 
\rightarrow \infty$. \\
Property i) implies that there are not string-like spikes in the dynamics.
As a consequence of property ii), the spectrum of the bosonic regularized
Hamiltonian consists of a discrete set of eigenvalues with 
finite multiplicity, see for instance \cite{Reed+Simon}. 
An explicit proof for the $g=1$ case was given in \cite{LB+MG+IM+AR}.

The fermionic sector of the Supermembrane Hamiltonian in the light 
cone gauge is 
\begin{equation}
  \int\limits_{{\cal M}} d^2\sigma \sqrt{w} \left( -\bar\theta \Gamma_-
  \Gamma_r {\cal D}_r \theta +\bar\theta \Gamma_- \Gamma_m {X^m,\theta}
  \right)\;.
\end{equation}
The Gauss constraint (\ref{Gauss}) acquires a fermionic term
\begin{equation}
  \{\theta\Gamma_-,\theta\}\;.
\end{equation}
In the regularized version of the supersymmetric Hamiltonian the 
fermionic\hyphenation{fer-mio-nic} fields are represented in terms of matrices satisfying the 
standard anticomutation relation \cite{LB+MG+IM+AR}. Therefore we deal in the
supersymmetric Hamiltonian with a matrix potential,
the bosonic potential appears only in the diagonal:
\begin{equation}
  V= V_B\mathbb{I} +V_F\;.
\end{equation}
$V_B$ is asymptotically greater or equal to the square norm in 
the configuration space while the fermionic matrix potential 
is linear on the configuration space variables. This property implies that 
the supersymmetric Hamiltonian has discrete spectrum with finite multiplicity. For two independent proofs of this statement see \cite{LB+MG+AR} and \cite{LB+AR}.

\section{Conclusions}
We constructed the Hamiltonian of the $D=11$ Supermembrane wrapped on a calibrated submanifold of the target space. The calibrated submanifold is defined by a minimal immersion of the supermembrane into the compact sector of the target space corresponding to a torus of dimension $2g$, where $g$ is the genus 
of the immersed base manifold. Thus we have extended the previous results 
of \cite{LB+MG+IM+AR,LB+MG+AR} where 
the target space was restricted to $S^1\times S^1$. The wrapping is 
a consequence of a topological condition on the space of configurations, which defines de $D=11$ Supermembrane with topological restrictions. The minimal immersion plays a central role in the construction. It allows to reformulate the Hamiltonian of the Supermembrane as  
a symplectic Yang-Mills one. We prove using properties of the 
symplectic Yang-Mills theory that the spectrum of 
its regularized Hamiltonian is discrete with finite multiplicity. 
The explicit construction for the case in which the base manifold is a genus $g=2$  compact Riemann surfaces shows that this Supermembrane, with the topological restriction, results as the intersection of two Supermembranes with their corresponding topological restrictions with base manifolds $g=1$ compact surfaces. The topological conditions work  in such a way that the gluing of the $g=1$ surfaces to give rise to a $g=2$ surface also induces an intersection of the corresponding membranes. The $D=11$ Supermembrane with topological restrictions over higher genus base manifolds are then the result of the intersection of Supermembranes over lower genus surfaces. The discretness of the spectrum is preserved. We also showed several properties of the minimal immersions, in particular that they are mininal configurations of D$p$-branes with K\"ahler base manifold.

\section{Acknowledgements}
The authors would to thank to A. Font and M. P. Garc\'{\i}a del Moral 
for useful comments and discusions. \\

\noindent {\Large {\bf Appendix. Holomorphic immersions and symplectic gauge\hyphenation{gau-ge} theories}} \\

\noindent In this appendix we list the mean definitions of symplectic Yang-Mills theories and describe how a holomorphic immersion can be used to translate symplectic structures between manifolds.

Let $\hat X^r$ denote both the real and imaginary parts of the components of a holomorphic immersion of a K\"ahler surface ${\cal M}$ into a complex flat torus $T$. Since the target space is flat and the base manifold is bidimensional the holomorphic immersion is also a minimal immersion of the K\"ahler surface into $T$. The almost complex structure, K\"ahler form and K\"ahler metric of ${\cal M}$ are denoted by $J^\mu{}_\nu$, $\Omega_{\mu\nu}$ and $K_{\mu\nu}$ and the analogous objects of $T$ are denoted by $j^r{}_s$, $\omega_{rs}$ and $g_{rs}$. 

The condition of $\hat X^r$ of being holomorphic can be covariantly stated as 
\begin{equation} 
	\partial_\mu\hat X^r J^\mu{}_\nu = j^r{}_s \partial_\nu\hat X^s\;.
\end{equation}  
The holomorphic immersion pulls the K\"ahler form and
K\"ahler metric of $T$ back to ${\cal M}$
\begin{eqnarray} 
&&	\Omega_{\mu\nu} = \partial_\mu\hat X^r \partial_\nu\hat X^s \omega_{rs} \\
&& 	K_{\mu\nu}= \partial_\mu\hat X^r \partial_\nu\hat X^s g_{rs}\;. 
\end{eqnarray} 

We may consider the following objects 
\begin{equation}
	e^\mu{}_r \equiv -\Omega^{\mu\nu} \partial_\nu \hat X^s g_{sr} 
\end{equation} 
they verify 
\barr 
&& 	\Omega^{\mu\nu}= e^\mu{}_r e^\nu{}_s \omega^{rs} \\ 
&& 	K^{\mu\nu}= e^\mu{}_r e^\nu{}_s g^{rs} 
\earr 
where $\omega^{rs}$ is the inverse of $\omega_{rs}$. In the case of $T$ is a bidimensional torus the holomorphic immersion is a bijective map and $\partial_\mu\hat X^r$ plays the role of zweibein over ${\cal M}$.

We may introduce the following derivatives 
\begin{equation} 
	D_r \equiv e^\mu{}_r \nabla_\mu\;.
\end{equation}
where $\nabla_\mu$ is the Levi-Civita connection of the K\"ahler metric over ${\cal M}$. The K\"ahler form of ${\cal M}$ is also covariantly constant respect to $\nabla_\mu$.

The K\"ahler form $\Omega$ provides a symplectic structure for ${\cal M}$,
analogously $\omega$ does the same on $T$. We define the symplectic bracket
on ${\cal M}$ 
\begin{equation}\label{bracket} 
	\{\phi,\varphi\} = \Omega^{\mu\nu}\nabla_\mu\phi\nabla_\nu\varphi
\end{equation} 
where $\phi$ and $\varphi$ represent generic tensors or pseudotensor over ${\cal M}$.
This bracket is antisymmetric and satisfies the Jacobi identity since the K\"ahler
form is closed. 

The above bracket may be rewritten as 
\begin{equation} \label{NewPB}
	\{\phi, \varphi\}= \omega^{rs} D_r\phi D_s\varphi\;. 
\end{equation} We
remark that this relation is a non trivial property of the holomorphic map
$\hat X$. Therefore we see that the symplectic brackets of the base manifold ${\cal M}$ and the ones of its image on $T$ throughout $\hat X^r$ are compatible via this map.

Now we consider a symplectic connection 
\begin{equation} 
	{\cal D}_r= D_r+\{{\cal A}_r,\:\:\} 
\end{equation} 
which implement the symplectic gauge
symmetry. The case of interest for us is when ${\cal A}_r$ are singled
valued smooth functions over all ${\cal M}$. Consequently, we define the
curvature 
\begin{equation} 
	{\cal F}_{rs}=D_r{\cal A}_s- D_s{\cal A}_r+\{{\cal
	A}_r,{\cal A}_s\}\;, 
\end{equation} 
such that 
\begin{equation}
	\parent{{\cal D}_r{\cal D}_s- {\cal D}_s{\cal D}_r}
	\phi=\{{\cal F}_{rs},\phi\}\;. 
\end{equation} 
The symplectic gauge transformation is defined as 
\begin{equation}\label{deltaA} 
	\delta_\xi {\cal A}_r = {\cal D}_r\xi\;, 
\end{equation} 
where $\xi$ is an infinitesimal parameter. For
the curvature we obtain 
\begin{equation}\label{deltaF} 
	\delta_\xi {\cal F}_{rs}=\{
	\xi, {\cal F}_{rs} \}. 
\end{equation} 

The symplectic Yang-Mills action is 
\begin{equation} 
	\int\limits_{{\cal M}} {\cal F}_{rs} {\cal F}^{rs} \Omega \;.
\end{equation} 
This action can be alternatively formulated over the image of 
${\cal M}$ in $T$ under the holomorphic map $\hat X$, which we denote by ${\cal M}'$, as
\begin{equation} 
	\int\limits_{{\cal M}'} {\cal F}_{rs} {\cal F}^{rs} \omega_{tu} 
	d\hat X^t \wedge d\hat X^u \;.
\end{equation} 
The invariance of the symplectic Yang-Mills action
under \ec{deltaA} follows from the identity 
\begin{equation} 
	\int\limits_{{\cal M}} \Omega f\{\xi,f \}=0
\end{equation} 
which ensures the invariance of the quadratic functional
$\left<f^2\right>$ under the homogeneous transformation 
\begin{equation}\label{Homogeneous} 
	\delta_\xi f=\{\xi,f \}\;. 
\end{equation}

Finally we notice that for the minimal immersion $\hat X^r$ into a bidimensional torus
we have
\begin{equation}
  \{\hat X^r, \hat X^s\} = \epsilon^{rs} n
\end{equation}
which is analogous to the commutator of coordinates on the noncommutative
torus
\begin{equation}
  [ \hat X^r, \hat X^s ] = \epsilon^{rs} \theta.
\end{equation}
For that reason the symplectic gauge theories have also been called
symplectic noncommutative gauge theories. The symplectic two form over
${\cal M}$ may be locally mapped to a constant one, via Darboux theorem.
On the torus the symplectic two form may be globally expressed as a
constant antisymmetric tensor. The noncommutative Yang-Mills theories
constructed with constant symplectic forms as well as its description in
terms of D-branes over constant antisymmetric field has been extensively
analyzed in the literature.


\begin{thebibliography}{99}
\bibitem{Bergshoeff}{E. Bergshoeff, E. Sezgin, P. K. Townsend,
  Phys. Lett. B 189 (1987) 75.}

\bibitem{LB+MG+IM+AR}{L. Boulton, M. P. Garc\'{\i}a del Moral,
  I. Martin, A. Restuccia, Class. Quantum Grav. 19 (2002) 2951, hep-th/0109153.}
  
\bibitem{LB+MG+AR}{L. Boulton, M. P. Garc\'{\i}a del Moral,  A. 	Restuccia, Nucl. Phys. B 671 (2003) 343, hep-th/0211047.}

\bibitem{Duff}{M. J. Duff, T. Inami, C. N. Pope, E. Sezgin, K. Stelle,
  Nucl. Phys. B 297 (1988) 515.}
  
\bibitem{IM+AR+RT}{I. Martin, A. Restuccia, R. Torrealba,
  Nucl. Phys. B 521 (1998) 117, hep-th/9706090.}  
  
\bibitem{IM+JO+AR}{I. Martin, J. Ovalle, A. Restuccia, Phys. Rev. D
  		64 (2001) 046001, hep-th/0101236.}
  		
\bibitem{MG+AR}{M. P. Garc\'{\i}a del Moral, A. Restuccia, Phys. Rev. D 66 (2002) 045023, hep-th/0103261.}
  
\bibitem{MPG}{M. P. Garc\'{\i}a del Moral, A new mechanism of K\"ahler moduli stabilization in Type IIB Theory, hep-th/0506116.}

\bibitem{JC+MPG+FQ+AU}{J.F.G. Cascales, M.P. Garc\'{\i}a del Moral, F. Quevedo, A.M. Uranga, JHEP 0402 (2004) 031, hep-th/0312051.}

\bibitem{deWitt+Luscher+Nicolai}{B. de Wit, M. L\"uscher,  H. Nicolai, Nucl. Phys. B 320 (1989) 135.}

\bibitem{deWit+M+N}{B. de Wit, U. Marquard, H. Nicolai,
  Commun. Math. Phys. 128 (1990) 39.}

\bibitem{deWitt+Peeters+Plefka}{B. de Wit, K. Peeters, J. C. Plefka,
  Nucl. Phys. Proc. Suppl. 62 (1998) 405, hep-th/9707261.}
  
\bibitem{Harvey+Lawson}{R. Harvey, H. B. Lawson, Acta Math.
     148 (1982) 47.}

\bibitem{Harvey}{R. Harvey, Perspect. Math. 9 (1990).}

\bibitem{Cascales+Uranga}{J. Cascales, A. Uranga, JHEP 0411 (2004) 083, hep-th/0407132.}

\bibitem{Smith}{Douglas J. Smith, Class. Quantum Grav. 20 (2003)
     R233, hep-th/0210157.}

\bibitem{B+R}{J. Bellor\'{\i}n, A. Restuccia, Phys. Rev. D
			64 (2001) 106003, hep-th/0007066.}
			
\bibitem{Trautman}{A. Trautman, Int. J. Theor. Phys. 16	(1977) 561.}

\bibitem{Leigh}{R. G. Leigh, Mod. Phys. Lett. A4 (1989) 2767.}

\bibitem{Matsushima}{Y. Matsushima, J. Diff. Geom. 9 (1974) 309.}

\bibitem{Nagano}{T. Nagano, B. Smyth, Comment. Math. Helvetici 50
			(1975) 249.}
  
\bibitem{LB+AR}{L. Boulton, A. Restuccia, The Heat Kernel of the compactified $D=11$ Supermembrane with non-trivial winding, hep-th/0405216, to be published in Nucl. Phys. B.}

\bibitem{deWitt}{B. de Wit, J. Hoppe, H. Nicolai, Nucl. Phys. B
			305 (1988) 545.}
  
\bibitem{NH+AR+JS}{N. Hatcher, A. Restuccia, J. Stephany, The $D=11$ Supermembrane wrapped on a two cycle and the KKB Superparticle in $D=9$, hep-th/0506042.}
			
\bibitem{Hoppe}{J. Hoppe, MIT PhD-Thesis (1982).}

\bibitem{Reed+Simon}{M. Reed, B. Simon, Methods of Modern Mathematical Physics, v. 4 (Academic Press, New York, 1978).}

\end{thebibliography}
\end{document}